%% file: main.tex
\documentclass[11pt,letterpaper]{article}

\pdfoutput=1

\usepackage{times}
\usepackage{relsize}
\usepackage{microtype}
\usepackage{cite}
\usepackage{xspace}
\usepackage{xcolor}
\usepackage{color,colortbl}
\usepackage{subfigure}
\usepackage{alltt}
\usepackage{url}
\usepackage{comment}
\usepackage{graphicx}
\usepackage{relsize}
\usepackage{paralist}
\usepackage{wrapfig}
\usepackage{titlesec}
\usepackage{epsfig}
\usepackage{geometry}
\usepackage{tikz}
\usepackage[normalem]{ulem}
\usepackage{titlesec}
\usepackage{booktabs}
\usepackage{wrapfig}
\usepackage{multirow}
\usepackage{pifont}
\usepackage{wrapfig}
\usepackage{soul}
\usepackage{enumitem}
\usepackage{comment}
\usepackage[some]{background}

\titlespacing{\section}{0pc}{1pc}{0.2pc}
\titlespacing{\subsection}{0pc}{1pc}{0.2pc}
\titlespacing{\subsubsection}{0pc}{1pc}{0.2pc}
\titleformat*{\section}{\normalsize\bfseries}
\titleformat*{\subsection}{\normalsize\bfseries}
\titleformat*{\subsubsection}{\normalsize\bfseries}

\definecolor{lightblue}{rgb}{.90,.95,1}
\definecolor{lightgreen}{rgb}{.90,1,.95}
\definecolor{lightgray}{gray}{.9}
\definecolor{lightorange}{rgb}{1,.85,.85}
\definecolor{lightpink}{rgb}{1,.85,.95}
\definecolor{lightcyan}{rgb}{.75,.9,.9}
\definecolor{lightpurple}{rgb}{.9,.83,.93}

\usepackage[font=small,labelfont=bf]{caption}
\usepackage{supertabular}

\iffalse 
\newcommand{\assign}[1]{\textcolor{blue}{TO-DO:} \textcolor{magenta}{#1}}

\def\add{\textcolor{blue}}
\def\comment{\textcolor{red}}
\else
\newcommand{\add}[1]{}
\newcommand{\assign}[1]{}
\renewcommand{\comment}[1]{}

\fi

\newcommand{\projectname}{Cybercosm\xspace}
\newcommand{\platform}{transvisor\xspace}

\newcommand{\lstore}{LStore\xspace}

\begin{document}

\pagestyle{empty}
\thispagestyle{empty}



\setcounter{page}{1}
\pagestyle{plain}




\input{cover-page.tex}

\input{sec-problem-statement}

\input{sec-Q_and_A}
\input{sec-proposed-research}
\input{sec-experimental-systems}

\color{blue}

\color{black}

\clearpage

\setcounter{page}{1}
\pagenumbering{Roman}
\bibliographystyle{IEEEtran}
\bibliography{expeditions, references}

\newpage
\setcounter{page}{1}

\newpage
\setcounter{page}{1}



\end{document}

%% file: cover-page.tex
\definecolor{titlepagecolor}{cmyk}{.8,0, 0.6, 0}

\DeclareFixedFont{\bigsf}{T1}{phv}{b}{n}{1.0cm}

\backgroundsetup{
	scale=1,
	angle=0,
	opacity=1,
	contents={\begin{tikzpicture}[remember picture,overlay]
			\path [fill=titlepagecolor] (-0.5\paperwidth,5) rectangle (0.5\paperwidth,10);  
	\end{tikzpicture}}
}
\begin{titlepage}
		\makeatletter                       
		\def\printauthor{%
			{\small \@author}}              
		\makeatother
		\author{%
			Mark Asch \\
			LAMFA, UPJV \\
			\texttt{mark.asch@u-picardie.fr}\\[1mm]
			Micah Beck \\
			EECS, UTK \\
			\texttt{mbeck@utk.edu} \\[1mm]
			Fran\c cois Bodin \\
			IRISA, Rennes \\
			\texttt{francois.bodin@irisa.fr} \\[1mm]
			Terry Moore \\
			ICL, UTK \\
			\texttt{tmoore@icl.utk.edu} \\[1mm]
			Martin Swany \\
            ISE, IU \\
            \texttt{swany@indiana.edu} \\[1mm]
			Michela Taufer \\
			EECS, UTK \\
			\texttt{taufer@gmail.com} \\[1mm]
			Jean-Pierre Vilotte \\
		    IPGP, Paris \\
			\texttt{vilotte@ipgp.fr}
		}
		\BgThispage
		\newgeometry{left=1cm,right=4cm}
		\vspace*{1cm}
		\noindent
		\textcolor{white}{\bigsf Cybercosm: \\ \\ New Foundations for a Converged Science Data Ecosystem}
		\vspace*{1.0cm}\par
		\noindent
		\begin{minipage}{0.35\linewidth}
			\begin{flushright}
				\printauthor
			\end{flushright}
		\end{minipage} \hspace{15pt}
		\begin{minipage}{0.02\linewidth}
			\rule{1pt}{300pt}
		\end{minipage} \hspace{-10pt}
		\begin{minipage}{0.7\linewidth}
			\vspace{10pt}
			\begin{abstract} 
			Scientific communities naturally tend to organize around data ecosystems created by the combination of their observational devices, their data repositories, and the workflows essential to carry their research from observation to discovery. However, these legacy data ecosystems are now breaking down under the pressure of the exponential growth in the volume and velocity of these workflows, which are further complicated by the need to integrate the highly data intensive methods of the Artificial Intelligence (AI) revolution. Enabling ground breaking science that makes full use of this new, data saturated research environment will require distributed systems that support dramatically improved resource sharing, workflow portability and composability, and data ecosystem convergence.
			
			The \emph{\projectname} vision presented in this manifesto describes a radically different approach to the architecture of distributed systems for data-intensive science and its application workflows. As opposed to traditional models that restrict interoperability by hiving off storage, networking, and computing resources in separate technology silos, \projectname defines a minimally sufficient hypervisor as a spanning layer for its “data plane” that virtualizes and converges the local resources of the system’s nodes in a fully interoperable manner. By building on a common, universal interface into which the problems that infect today's data-intensive workflows can be decomposed and attacked, \projectname aims to support scalable, portable and composable workflows that span and merge the distributed data ecosystems that characterize leading edge research communities today.

			\end{abstract}
		\end{minipage}
	\vfill
	\centering 
	\vspace{15pt}
	{\scshape April, 2021} \\ \vspace{5pt}
	{\large ---MANIFESTO---} \par

	\end{titlepage}
	\restoregeometry

%% file: sec-problem-statement.tex

\section{Overarching Vision and Goals of the Science \projectname}
\label{sec:vision}

Since all modern science is based on observation, it is no surprise that, at every scale from the campus to the planet, scientists organize their research communities around the \textit{data ecosystems} created by the combination of their observational devices (e.g., instruments, sensors, simulations), their data repositories, and their computational workflows. But while research communities rely fundamentally on such data ecosystems, and computer scientists aspire to create them at planetary scale~\cite{Cohen2018-zv}, the foundations of these systems are increasingly in disarray. 

Both the volume of data and the velocity of its production are growing exponentially, due largely to the explosive improvements in and proliferation of data generators, ranging from the Internet of Things (IoT) and commercial clouds to scientific, engineering, medical, military, and industrial equipment of all kinds. Under the pressure of this data tsunami, the legacy data ecosystem---in which data is repeatedly put into and taken out of traditional technology silos for computing, networking, and storage---has become progressively more complex, expensive, and ad hoc. This systemic unraveling is made even more problematic through the data intensive workflows of the revolution in Artificial Intelligence (AI), which is radically transforming methods and techniques across every field of inquiry. Finally, the growing desire to intensify interdisciplinary cooperation and integration is routinely obstructed by the fact that the data ecosystems of different communities are hived off from one another by the same 
siloing that balkanizes research workflows. 

\begin{wrapfigure}[18]{r}{.75\textwidth}
\includegraphics[width=4.5in]{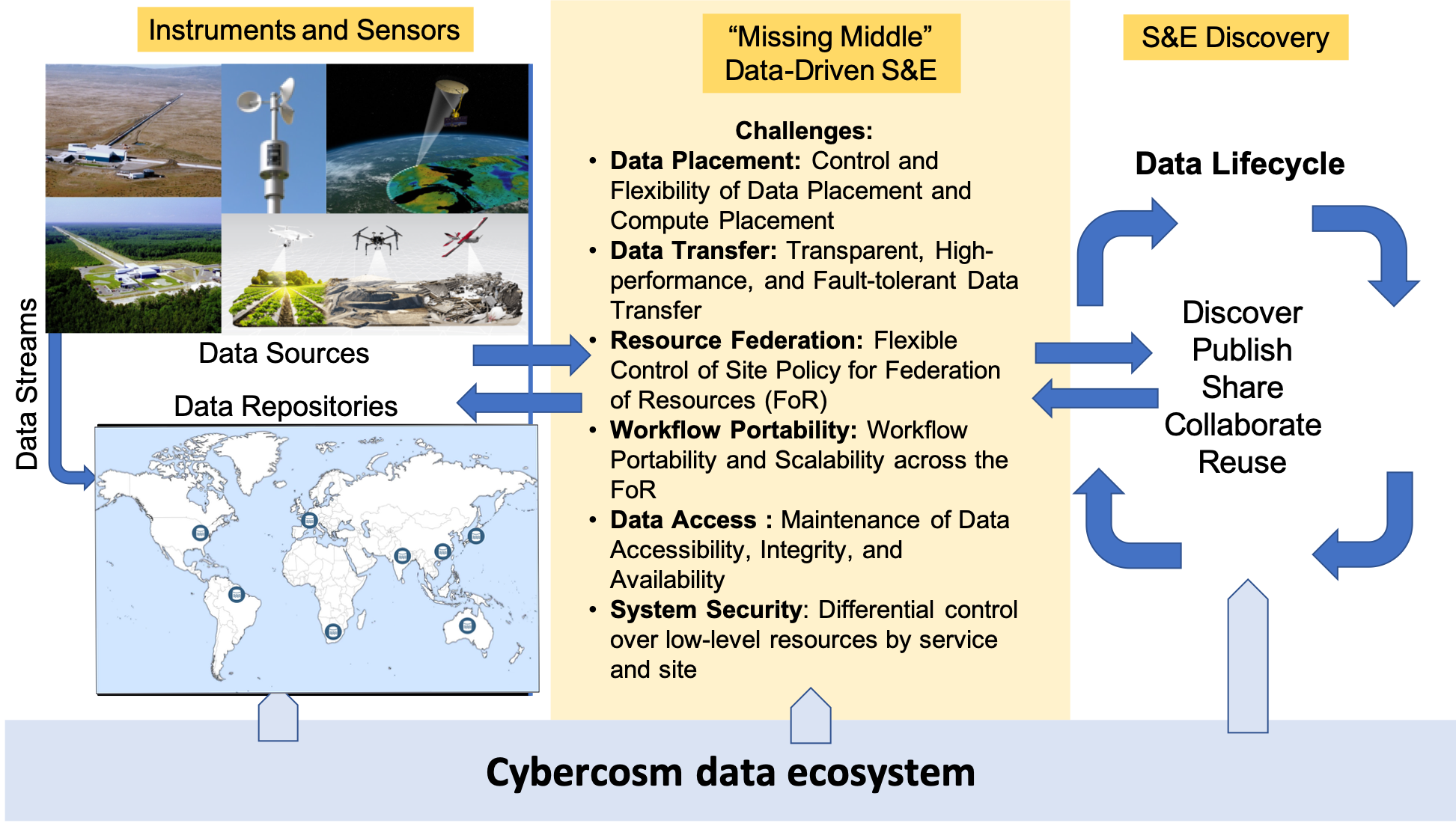}
\caption{The problem of 
creating a planet-scale data ecosystem for science is the problem of discovering 
a basic foundation for  the ``missing middle'' of science's cyberinfrastructure with planet-scale
deployment scalability (Adapted from~\cite{Parashar2020transforming}).
\label{f_middle}}
\end{wrapfigure}

\begin{wrapfigure}{l}{.40\linewidth}
\includegraphics[width=1.\linewidth]{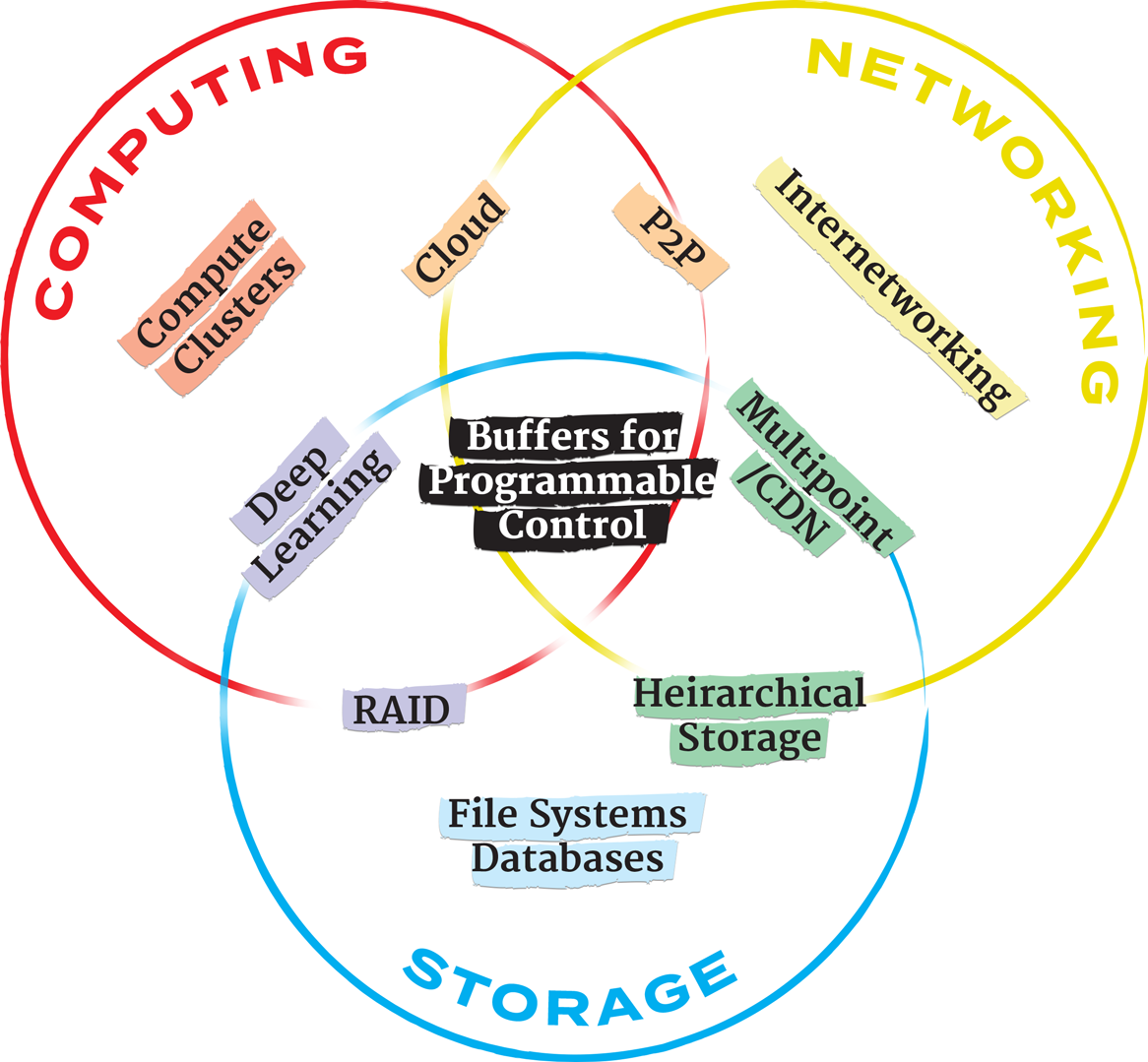}
\caption{Buffers, i.e. mechanisms for generic data persistence, lie at the intersection of the system silos for storage, networking, and computing. The spanning layer of the \projectname exposes a generic interface to buffers.}
\vspace*{-3ex}
\label{buffers}
\end{wrapfigure}

\begin{wrapfigure}{r}{0.5\linewidth}
\vspace*{-1ex}
\centering 
\includegraphics[width=0.8\linewidth]{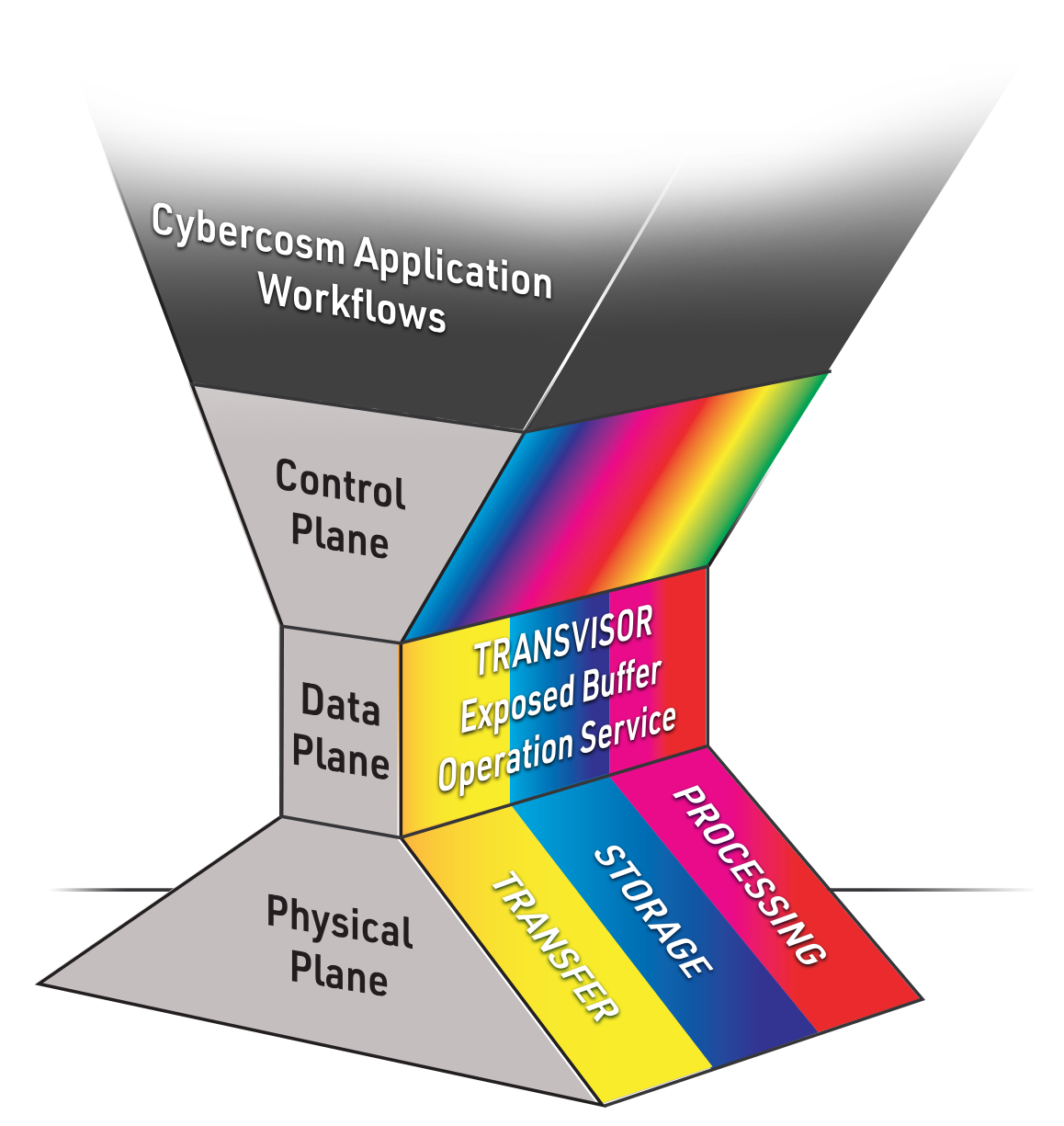}
\caption{The \platform service virtualizes the physical resources of nodes in the data plane. Control plane services aggregate \platform operations to create complex abstractions for \projectname application and ecosystem workflows.}
\label{transvisor_system_design1}
\end{wrapfigure}

Constructing a revolutionary new foundation for data ecosystems and their workflows---the  \textbf{``missing middle''} (Figure~\ref{f_middle})---is a fundamental challenge confronting cyberinfrastructure research communities today. This new foundation must integrate the disparate functions of storage, networking, and computation, but without just re-bundling legacy interfaces in yet another unwieldy overlay. In order to create such a converged data ecosystem that can support the scientific community generally, which we call the ``\textit{Science Cybercosm},'' the cyberinfrastucture community needs to articulate, catalyze, and carry out the fundamental research agenda required to unify, and thereby revolutionize, the foundations on which tomorrow's dramatically more interoperable data ecosystems will be built. Our \projectname vision is based on \textit{an observation}, \textit{a principle}, and \textit{a strategy} 
that together
open up a transformative path forward for cyberinfrastructure research and development generally: 

\begin{compactitem}

    \item \textbf{An observation about the centrality of buffers}: Buffers, which are regions of a physical medium (e.g., memory, disk storage, tape) where data persists while waiting to be moved or transformed, are the common building blocks for all digital information systems (Figure~\ref{buffers}). Buffers are essential for state management and therefore required for programmable control. Distinctions between different "container" technologies that are aggregations of specialized  types of buffers (e.g., address space, file, network flow) can be expressed by policies applied to their management and the correlated requirements for the material medium. Thus, computation is implemented as the repeated transformation of a set of pages (and associated control state) that are managed collectively as an operating system task; files and flows are analogous aggregates for storage and networking. Adopting this analysis, the spanning layer of the \projectname exposes, as its basic unit of service, \textit{a common, generic interface to a broad class of buffers and operations expressed in terms of buffers, performed locally}. We call the mechanism that implements this interface the \textbf{\platform},  i.e., \emph{a minimally sufficient hypervisor that provides a converged virtualization of the \textbf{local} resources---buffer, processing, and transfer---of the individual nodes of a distributed system}. In this \textit{exposed buffer} paradigm, the \platform's primitive operations on buffers are aggregated to construct the familiar abstractions of process, file, and flow~\cite{beck2019interoperable}. But these same primitive operations can be used to compose entirely new abstractions that cross the boundaries of technology silos with hybrid or novel aggregates, opening up a vast new opportunity space for technological innovation. 
    
    \item \textbf{A principled approach to deployment scalability:} In layered protocol and software stacks, the function of a \textit{spanning layer} is to hide the detailed differences among various implementation technologies beneath it while presenting a uniform service interface to the applications above it~\cite{clark1995interoperation,messerschmitt2003software}. Thus, in our \projectname vision, the \platform is the spanning layer that forms the ``waist'' of an iconic ``hourglass'' system design, as in Figure~\ref{transvisor_system_design1}. Successful spanning layers (e.g., the Internet Protocol Suite, POSIX) create communities of interoperability that benefit from innovation in the technology substrate without being disrupted by it. Since the goal of the ~\projectname is to create an interoperable data ecosystem for the scientific community generally, its fundamental question is "How can we create a spanning layer that can achieve, through \emph{voluntary} adoption, near universal deployment, i.e., that maximizes \emph{deployment scalability}?" There are compelling arguments to show that this question can be answered by attending to the \emph{deployment scalability tradeoff}: ``There is an inherent tradeoff between the deployment scalability of a [service] specification and the degree to which that specification is [logically] weak, simple, general and resource limited" ~\cite{beck2019hourglass}. Guided by this principle, the \projectname's platform spanning layer seeks to maximize deployment scalability and thus the associated network effects and ``future-proofing'' that come with it.

    \item \textbf{The \projectname strategy:} Since the goal is to create a new foundation for distributed systems that facilitates resource sharing, workflow composability, and interoperable convergence among the data ecosystems of diverse scientific communities, the question of a dissemination or ``shaping strategy''~\cite{Pathways2018,hagel2017shaping} has to be faced. Over the past twenty five years, ``If we build it, they will come \ldots'' has been the implicit strategy of a series of major research and development efforts proposing new spanning layers with similar aspirations (e.g., Active Networking~\cite{Tennenhouse96towardsan}, the Grid~\cite{foster2001anatomy}, PlanetLab~\cite{peterson2006experiences}, GENI~\cite{peterson2007overview}). However, none of these examples were as successful in achieving the kind of deployment scalability within their respective application communities as the Internet and the Unix kernel interface exhibited in their application communities. The \platform layer of \projectname is designed specifically to succeed on this front. By building on a weak, simple, and generic spanning layer that provides pervasive, low-level interoperability and portability,  
    \projectname enables a shaping strategy that, we believe, will maximize the \textit{voluntary} participation of users and adopters (e.g., through facile service composition and creation, federated resource sharing, and network effects), while lowering or minimizing the risks (e.g., vendor/technology lock-in, obsolescence, lack of interoperability) that so often raise barriers to adoption and use.

\end{compactitem}

The \projectname architecture supports data intensive workflows that span data ecosystems by building on a universal service---the \platform service---into which all such problems can be decomposed and attacked. It carefully divides responsibilities between a control plane and a data plane in order to simplify system management and service deployment (Figures~\ref{transvisor_system_design1}~\&~\ref{f_planes})
. This strict separation enables control plane services to share the metadata that defines a file or process without physically moving the underlying data objects. 

In the \projectname system, the atom of state is a buffer in the data plane; the atom of transformation is an operation on such a buffer; and the atom of movement is transfer between buffers on the same or adjacent nodes. Workflows are constructed by control plane services that aggregate these atoms; data is generated into buffers or has been previously stored in them. Following ``instructions'' from the control plane, each data plane operation acts upon a set of co-located buffers, reading or writing as needed. Workflows are expressed as the application of operations on buffers and transfers between buffers, although their allocation and control 
may be determined dynamically at runtime. The scheduling problem is to map these operations onto the graph of available resources: \begin{inparaenum}[(1)]  \textit{\item In-transit} computation locates processors along paths where data may be transferred and applies operations when intersections occur, thereby moving the data to the computation;  \textit{\item In-situ} computation locates processors where data is stored and applies operations to stored data, bringing computation to the data;  \textit{\item In-locus} computation is not constrained by either of these paradigms but creates a data plane resource aggregation in which data transfers and operations can be intermixed in any pattern that the control plane can devise and manage.\end{inparaenum}

A defining feature of this novel architecture, one which simplifies system management and service deployment, is the rigorous separation it establishes between a control plane and a data plane (Figures~\ref{transvisor_system_design1}~\&~\ref{f_planes}), much as Software Defined Networking (SDN) does. Services such as globally routed packet delivery, process management, and file management are implemented in \projectname’s control plane, which takes responsibility for some of what the Internet delegates to the forwarding (i.e., data) plane and to the operating systems of network attached servers, computational and storage nodes. The \projectname data plane takes responsibility only for executing weak, simple, and generic operations (on buffers) that can be limited in many dimensions (e.g., duration, input size, or memory utilization). In this sense, transvisor ``packetizes’’ data persistence and processing the way that packet networking does for data transmission. It uses a unique function-as-a-service model that integrates processing with data persistence and data transfer; the ``network functional unit’’ (NFU) operations that do the work are semi-statically defined operations bound in a global namespace analogous to Internet port assignment (i.e., they are not mobile code). These operations act on local buffers, are limited in their use of all resources, and are best effort.

For understandable reasons, the radical nature of \projectname’s architectural vision is likely to elicit various questions from readers of this white paper. Although endeavoring to anticipate and answer many of these queries would quickly exceed the scope of this declaration of vision and purpose, responding to a small sample of them might help contextualize our enthusiasm for this unique point of view. Accordingly, below we offer, in turn, brief responses to two important questions that seem likely to arise: First (Sec.~\ref{alternatives}), How does \projectname compare and relate to major ongoing efforts to build a new data ecosystem for science? Second (Sec.~\ref{security}), Does \projectname have any better potential for innovations that might help solve the chronic (and growing) security and privacy problems that assail Internet-based systems today?

%% file: sec-Q_and_A.tex

\subsection{Cybercosm and Alternatives: Some Context}
\label{alternatives}

The unprecedented waves of data that are now flooding research communities have naturally spurred various efforts to design and deploy distributed data ecosystem platforms that can help the scientific community cope with this flood of data. Examples from the traditional HPC cyberinfrastructure community include the Data Lakes initiative of the Worldwide LHC Computing Grid~\cite{bird2019architecture}, Open Storage Network~\cite{OSN}, National Research Platform~\cite{smarr2018pacific}, and Helmholtz Federated IT Services~\cite{suarezdeveloping}.  The existence of such alternatives raises the question of how the \projectname vision presented here compares and contrasts with these efforts. But in this white paper we do not give a detailed, comparative analysis of alternative systems for two reasons, one relating to the scope of the comparison set, the other to the question of the appropriate perspective from which to evaluate such alternatives. 

On the question of scope, it seems clear that any evaluation of alternative directions for data ecosystem infrastructure must consider the option of utilizing the utterly massive, and massively funded, commercial cloud infrastructures  (e.g., Amazon, Microsoft, Google, Alibaba, IBM, etc.) now vying to dominate the Infrastructure as a Service (IaaS) market. The creation of  the NSF funded Cloud Bank~\cite{cloudbank} and reports of national academies~\cite{NAP21886} show that, in the US at least, there is already significant movement in that direction. But it is also widely recognized that outsourcing critical research infrastructure to an IaaS provider presents not only significant opportunities and but also risks, e.g.  managing cloud, non-cloud, and inter-cloud data workflow topologies.  Putting such commercial IaaS offerings into comparison with \projectname and other publicly funded data infrastructure designs mixes technical questions with a host of economic, social, and political issues that are well beyond the scope of this manifesto.

On the question of perspective, the point of view traditionally adopted in the HPC-centric world for evaluating platform architectures typically sees peak performance as its primary goal, with relevant metrics established for each of the different ICT silos— storage, networking, and computing. From this perspective, the challenge in constructing a large scale, geo-distributed platform for data intensive science is to provide an overlay interface (or set of such) that makes it possible to compose the services of these performance-optimized silos into sustainable workflows that support the needs of specific research communities. While this perspective acknowledges that the interoperability/portability of workflows across geo-distributed data ecosystems is an important goal for scientific cyberinfrastructure, successful workflow exemplars on this conventional path have tended to be highly customized and ecosystem/infrastructure specific.

By contrast, the goal of \projectname is to create \textit{an interoperable data ecosystem for the scientific community generally}.  We take this as our goal in response to the fact that modern science is a collective exercise in social cooperation that is carried out by communities of inquiry which strive to work together across geographical, organizational, and generational boundaries. The next generation data ecosystem platform on which science builds needs to facilitate collaboration and resource sharing in a sustainable way across all these types of boundaries and in all their variety. We believe that this implies, in turn, that the \projectname platform needs to be designed from the bottom up with a common spanning layer that maximizes generality and interoperability and portability. By applying the deployment scalability tradeoff to the design of the \platform’s exposed buffer service, we hope to enable the largest possible number of implementations, and thereby facilitate voluntary adoption and use of the \projectname’s \platform spanning layer across the entire community. 

But since this exposed buffer architecture seeks to maximize generality and interoperability first, building up from a converged, low-level interface to the local resources of data plane nodes, and only then to address the question of how performance can be achieved in more specialized applications, two obious questions arise: 

\begin{compactenum}
\item What about performance, especially in regard to those legacy, performance-optimized services that the current siloed infrastructure addresses so well.

\item Can this approach enable service creators to compose novel abstractions that dissolve the ossified walls of the current technology silos in order to enable the kind of distributed service innovation that the emerging world of IoT, cyber-physical systems, and the ``Computing Continuum’’~\cite{beckman2020harnessing} seems to call for?
\end{compactenum}

Providing satisfactory answers to these questions would require us to go beyond the objectives of this white paper. However, it is worth pointing out that the strategic issue behind (2) throws into sharp relief the contrast between the Cybercosm perspective and the ``No reinventing wheels!'' perspective of the conventional path. According to the latter, the strategy for addressing an increasingly data saturated future, full of many difficult and unknown challenges, is to accept the relatively specialized and highly optimized technology silos as currently entrenched, and then to adapt to the challenges as they arise by building and integrating services using those silos. According to the former, the strategy is to lay down a new converged platform substrate as a spanning layer that maximizes generality and interoperability, and then work to rebuild and re-optimize familiar services and innovate new ones as goals and special circumstances require. The discussion of security alternatives below provides a partial illustration of this contrast.

\subsection{Potential for \projectname Innovation in Security and Privacy}
\label{security}

Since the effort to develop next generation data ecosystems for science is taking place against the backdrop of a rising tide of security and privacy debacles and disasters now enveloping Internet-based cyberinfrastructure, proposed design ideas will inevitably be evaluated, at least in part, by their capacity to change this perilous dynamic in a positive direction. It is therefore important to recognize that a major source of this dynamic derives from one of the factors that made the Internet grow and spread so quickly, namely, from the choice of \textit{end-to-end datagram delivery} as the universal service that defines its common ground. This service shares resources across the wide area network through the use of routers and the exchange of routes (e.g., through BGP). As the history of the Internet shows, by giving every node in the network access to the receive buffers in every other network-connected node, this choice was a powerful enabler of rapid client/server and peer-to-peer application development. 

But giving such power to senders means saddling receivers with corresponding levels of vulnerability. Once a route has been advertised to a peer, a router has little basis on which to make admission decisions, and transit through the router is provided to all. Transit through a router implies that that router will allocate resources on the next router in the  path to the degree that it can, acting in effect as a full proxy for the sender. The service leaves it up to every network-connected node to defend itself from malicious or otherwise harmful packets that are automatically forwarded to it. Moreover, it provides no protection to the hardware and software mechanisms (e.g., firewalls) along the path from the network signal handlers to those receivers (in the operating system or in application processes). Given the way the world is, the result is frequent hacking or cyberwarfare attacks on those mechanisms.

By contrast, the transvisor service at the spanning layer of the \projectname is a buffer service that expresses all the work to be done in terms of operations on buffers of data performed \textit{locally}; the \textit{data plane} mechanism that implements this service creates a converged virtualization of the resources — buffer, processing, and transfer — that are \textit{local} to the system’s nodes. Securing the resources of a local node (and its adjacent peers) is an easier task because the mechanisms used do not need to scale to the entire community of network users, only to those with direct access to the node. Access to the resources on non-local nodes is accomplished through services implemented in \projectname’s \textit{control plane} using \textit{service-specific peering mechanisms and agreements that can be specialized to the community served}. 

For example, constructing a robust and scalable file distribution system is a complex problem that works by building up many layers, both within the control plane and at the application layer. An important aspect of this layering is that different resource and information sharing policies can be and typically will be applied at every level. The lowest level services in the control plane make use of the local resources of nodes via the transvisor interface, but they can also define the terms of connectivity. Although services are free to adopt common protocols to meet their needs, if a service chooses not to expose particular nodes to the kind of unmediated datagram forwarding service that defines the Internet, it can also hive them off from global routed traffic.

Of course merely enabling application-specific policy and restricted access does not automatically create secure resource enclaves.  What it does create is an environment in which the definition and adoption of protocols and services can proceed without the constant bombardment of malicious traffic from the entire worldwide network. So in the \projectname model, good fences make good neighbors, but they do not obviate the need for building up trustworthy relationships and the policies and mechanisms to implement them.

In summary, cybersecurity is a major issue for today's global and network-enabled scientific projects. The Cybercosm approach \emph{does not} impose a centralized, one size fits all cybersecurity policy, but provides a framework that factors the problem between administration of local nodes in the data plane and the freedom to adopt or reject potentially global services in the control plane.  Local ICT administrators can impose their own checks and balances on nodes in their domain; it is their responsibility to see that {\em each node} implements adequate cybersecurity measures for protection at the local level. Then, when control plane services span multiple localities, it is the responsibility of {\em each of them} to implement an appropriately secure notion of service-specific peering or federation; individual domains are free to adopt or reject control plane services, depending on what their user communities require. Some communities could even decide that promiscuous datagram forwarding is a bad idea. By designing from the bottom up with deployment scalability as a priority, \projectname can offer an approach to federation and peering through differentiated services that can adapt to the security requirements of different user communities.


%% file: sec-proposed-research.tex

\section{Enabling Data-Driven Science and Engineering Discovery}
\label{sec:vision}

We believe \projectname would free domain scientists who are pursuing bold research agendas from the constraints imposed by the legacy resource silos of administratively centralized data centers, whether at single site, such as a traditional HPC center, or in a global network of such sites, like a commercial cloud. 
Using \projectname and the \emph{shareable} ``missing middle'' resources it provides, domain researchers would have the flexibility to reconfigure and move their data-intensive research as their needs and priorities change. Since no off-the-shelf solutions for shareable ``missing middle'' infrastructure for data ecosystems offers this capability, domain researchers are often forced to custom-build their own solutions for their unique problems. But this status quo appears to be unsustainable in terms of the costs incurred --- money, time, and opportunity --- by research teams with specialized needs.
 
Below we present three different domain science use cases to show the variety of application level requirements that we expect \projectname to address. Table~\ref{sec:usecase_list} then summarizes a wider spectrum of other application domains that \projectname can support.

\subsection{Driving Use Case: Windows on the Universe}
\label{sec:usecase1}
\textbf{Background}. A never-before-seen vision of the cosmos is emerging from observations of photons, cosmic rays,
neutrinos, and gravitational waves. The observations are providing hints to answer fundamental
questions, such as the nature of dark matter, tests of general relativity, the extreme conditions in
neutron stars and their role in gamma ray bursts, the type of black holes in the universe, and the
sources of very-high-energy neutrinos and cosmic rays. Harnessing these observations to understand
the physical processes at work in the universe is the core mission of the Center for Relativistic
Astrophysics (CRA) at Georgia Tech. Success requires analysis of distributed data sets that
enables multi-messenger discovery in theoretical and experimental astrophysics~\cite{Abbott:2016blz}.

\textbf{Workflow Interoperability}. Leading astroparticle observatories that detect gamma rays (CTA, HAWC, and VERITAS), neutrinos (IceCube), and gravitational waves (the NSF Laser Interferometric Gravitational Wave Observatory, LIGO) generate data to be analyzed. Identifying a signal buried in either
noise (LIGO) or in a vast number of background events (astroparticle) is a challenge. Billions of
simulated signal and background events are required to train background rejection methods. Current workflows are dictated by complex interplays of data sources, real and perceived costs of data transport,
technology and organizational limits, requirements of individual software packages, and the availability of particular pieces of hardware. Access and use of data for modeling requires user accounts, resource provisioning, and scheduling, all of which is further complicated by frequent personnel changes in Ph.D. students and postdocs. Some key components of the workflow are restricted by geographic locations or the availability of resources. For example, components of the workflow and various software packages are often suited for a variety of systems such as large memory single node, multicores, large multicore clusters, and GPU clusters. 

\textbf{Portable and Composable Workflows}. The \projectname architecture will support the logistics of such
digital workflows by building a universal service into which all such
problems can be decomposed and tackled. In the \projectname's \platform system, the simulations and data analytic
tools can be phrased in terms of operations on buffers of varying sizes and duration. Computation can naturally take place at
the data source, or based on data migration to a central computing resource, or anywhere in between,
all implemented by the \platform, guided by the control plane and oblivious to the user. Each operation acts upon a set of co-located
buffers, reading or writing as needed. Buffers for event detection may involve small subsets of data that can be loaded onto the memory of a single node, or multiple nodes, to up to hundreds of terabytes creating a large demand to write to disk and requiring high I/O throughput and data locality\cite{bhat2004high}. These issues are
seamlessly handled by the Cybercosm stack. 


\subsection{Driving Use Case: Critical Societal Infrastructures}
\label{sec:usecase2}

\textbf{Background}. Critical societal infrastructures (such as power, energy, transportation and water) are complex interdependent systems which are vital to public life and national security. Failure of even a small part of these systems, whether caused by natural or humanmade disaster, can trigger widespread cascading failures impacting many other interdependent modules and disrupting the functionality of the entire system. As a vivid example, the massive power outage during the 2003 blackout in the northeast US cascaded to impact water-waste treatments, transportation, communication and food industries. To better manage the effects of such disruptive events, domain experts should be able to asses, in a comprehensive manner, the complex interdependencies and failure dynamics over these systems. This is challenging due to the multiple layers of multidimensional data (such as types of interdependencies, types of failures, and coupling) at various temporal and spatial scales. Much of this data is also dynamic and can change frequently. Several toolkits in the critical infrastructure systems space are being used by Department of Energy entities and utility companies~\cite{tabassum2019}. 

\textbf{Federated Heterogeneous Resources} The \projectname architecture will be beneficial to critical societal applications in multiple ways, due to its support for in-situ, in-transit, and in-locus computation via location independent buffers. Firstly, it can help with creation and processing of large-scale heterogeneous networks denoting connections across multiple subsystems with data distributed over different geographical regions. Secondly, it can help in using ML and algorithmic techniques to model high-fidelity complex dynamics over these networks, which frequently requires both local neighborhood-based as well as non-local path-based operations (unlike much of the standard work in social networks). These frameworks help domain experts model "what-if" scenarios efficiently and identify possible emerging hotspots to take more informed restoration decisions. Finally, it can also help collect and analyze outage data coming from various monitoring devices and sensors both for real-time as well as retrospective batch-driven ML/AI analysis of sudden phase changes and anomalies over multiple time-scales to improve system resiliency.

\subsection{Driving Use Case: Arctic Ice-Sheet Modeling}
\label{sec:usecase3}
\textbf{Background}. The traditional workflows of ice sheet modeling are representative of other aspects of Earth system modeling. Observational data is produced by satellite missions or literal expeditions collecting ice cores, ground-based and aircraft-based observations, and other data. Large, complex models of the ice sheets alone or as part of larger Earth system models must be calibrated by this observational data to provide meaningful climate projections~\cite{Tezaur2015}. While the final product is necessarily integrative, the component data sets and computational transformations remain highly compartmentalized by fields of expertise. The agencies that curate expedition data perform their own data cleaning; component experts first integrate multiple data sets of the same observable, then smooth and interpolate them to standardized grids; system experts use regional models to integrate multiple observables into data sets that can be used to calibrate larger, higher-order models. This fractured chain of data transformations suffers from the same logistical issues as in virion modeling: inefficiency, friction, brittleness, and irreproducible results.

\textbf{Reproducibility of Science}. The \projectname architecture will perform all the data transformation steps within a single infrastructure
for greater resilience and reproducibility. This can be achieved without the explicit collaboration of all
domain scientists on one project; in keeping with services like Zenodo that encourage the use of software as
publishable, archivable, intellectual products, data set curators will be able to publish their Control Plane
transformations in addition to their data products. The same complexity that multiplies logistical difficulties in large computations like ice sheet models also multiplies sources of uncertainty in their projections; each transformation can have associated aleatoric, epistemic, and/or model inadequacy errors that uncertainty quantification should take into account. In the same way that the Cybercosm paradigm will facilitate resilience and reproducibility, it will also promote the composition of models of uncertainty (each again generated by domain experts). For example, the conductors of satellite missions have the best understanding of the precision of their instruments, the degree of correlation in the errors between data points, and of the atmospheric conditions at the time of measurement. All this information can be combined into a “noise model” accounted for within the Control Plane that can
generate realizations in-locus for ensemble modeling.

\subsection{Other Use Cases Across Broad Application Areas}
\label{sec:usecase_list}
In Table~\ref{tab:apps} below, we expand the discussion to a wider set of domains and applications than the three above, all targeted by \projectname.
\begin{table}[!h]
\small{
\tablefirsthead{\toprule Domains & Science and applications & Supported requirements \\ \midrule}
\tablelasttail{\\\bottomrule\\}
\bottomcaption{Domains, science and applications, and supported requirements. \label{tab:apps}}
\begin{supertabular}{p{5em} p{19em} p{19em}}
\small
Genomics
&
Perform alignment based database search for distributed delivery of services such as emerging pathogen detection or personalized genomics; requires analysis of high-throughput sequence data generated at laboratories/clinics worldwide.
&
Move analytics close to data generation; select resources with specific properties (e.g., large memory); define algorithms that are distributed and scalable.
\\
High Energy Physics
&
Store, distribute, and analyze data generated by the Large Hadron Collider (LHC); make data equally available to all partners, regardless of physical location.
&
Distribute storage and data; use ML techniques across data collection, processing, and analysis workflows to replace resource-intensive reconstruction and simulation algorithms.
\\
Radio Astronomy
&
Store, distribute, and analyze data generated by the Square Kilometer Array (SKA) among all partners of the geographically distributed SRC (SKA Regional Centers) network.
&
Provide mechanisms for a federation of regional SKA centers to uniformly present data and computing resources to the user. 
\\
Molecular Structures
&
Study large-scale, atomic-level molecules including binding affinities, folding, phase diagrams, rare events, protein engineering.
&
Move from offline to online analysis: support data reductions and adaptive simulations; stream data into analysis; integrate AI-driven analysis and simulations. 
\\
Materials properties
&
Study large-scale embodiments of the atomic-level structure of polymers in order to determine their physical, chemical, mechanical, and electronic properties; train learning algorithms to establish structure-property relationships.
&
Pre-compute  data analysis prior to arrival at facilities; share data and simulations across multiple sites; perform  analyses collectively at sites and reduce overall number of computing resources required.
\\ 
Precision Farming \& Soil Moisture Trends
&
Study soil moisture trends in precision farming by leveraging role of real-world satellite and local sensor data in soil moisture prediction simulations; support multiple orders of magnitude more data from satellite with global, coarse-grained resolution to local sensor data with local, fine-grained resolution
&
Meet strict real-time requirements of intense floating-point operations and I/O speed; fuse data from HPC-based high-accuracy simulations with real-world data; select future scenario based on simulations with quantified uncertainties.
\\
Smart Cities
&
Study traffic and emission trends in urban areas by collecting data from abundance of smart sensors/IoT devices; support fast-decision interactive services (real-time local decisions); process capabilities at the sensor side.
&
Build interoperability between edge devices and computing nodes; support efficient communication protocols and efficient data handling.      
\\
Wildfire Prediction \& Prevention 
&
Study wildfire behavior through modeling applications and WIFIRE cyberinfrastructure; collect fire perimeter from imagery captured by ground-based cameras, satellites, and aircraft.
&
Model dynamic-data driven workflows with continuous adjustment of fire modeling ensembles using observations; collect real observations for both workflow adjustments and perimeter propagation predictions.
\\  
\end{supertabular}
}
\normalsize
\end{table}

%% file: sec-experimental-systems.tex
\section{The \projectname Experimental System}
\label{sec:experimental-systems}

The software foundation on which \projectname's experimental infrastructure will be constructed is a product of a research program in Logistical Networking begun more than two decades ago, long before the data tsunami had overwhelmed the stateless model of networking and ``AI everywhere'' made data-centric infrastructure a necessity. Logistical Networking grew out of caching and server replication as an overlay solution to bringing generic state management to the Internet architecture. The model evolved, incorporating computation and mechanisms for dealing with latency between the control and data planes and supporting the internal functions of operating systems, file systems, and networks based on a generic abstraction of buffers. The task of justifying the adoption of a local model of buffers as the foundation of interoperability led to a formalization of the relationship between weakness of the spanning layer and deployment scalability using the tools of program logic. The artifacts of this program of research and deployment are now available to use as tools for building \projectname's experimental system and conducting new research on it. 

The lessons of the past 20 years have also led to new architectural insights and implementation strategies that can be used in \projectname's experimentation and deployment. \projectname provides not just a specific set of technological or analytical tools but a new approach to system architecture and interoperable cyberinfrastructure based on adopting more accurate abstractions of the infrastructure at a low level.

\begin{wrapfigure}[23]{r}{.55\textwidth}
\vspace*{-4ex}
\centering
\includegraphics[width=3.5in]{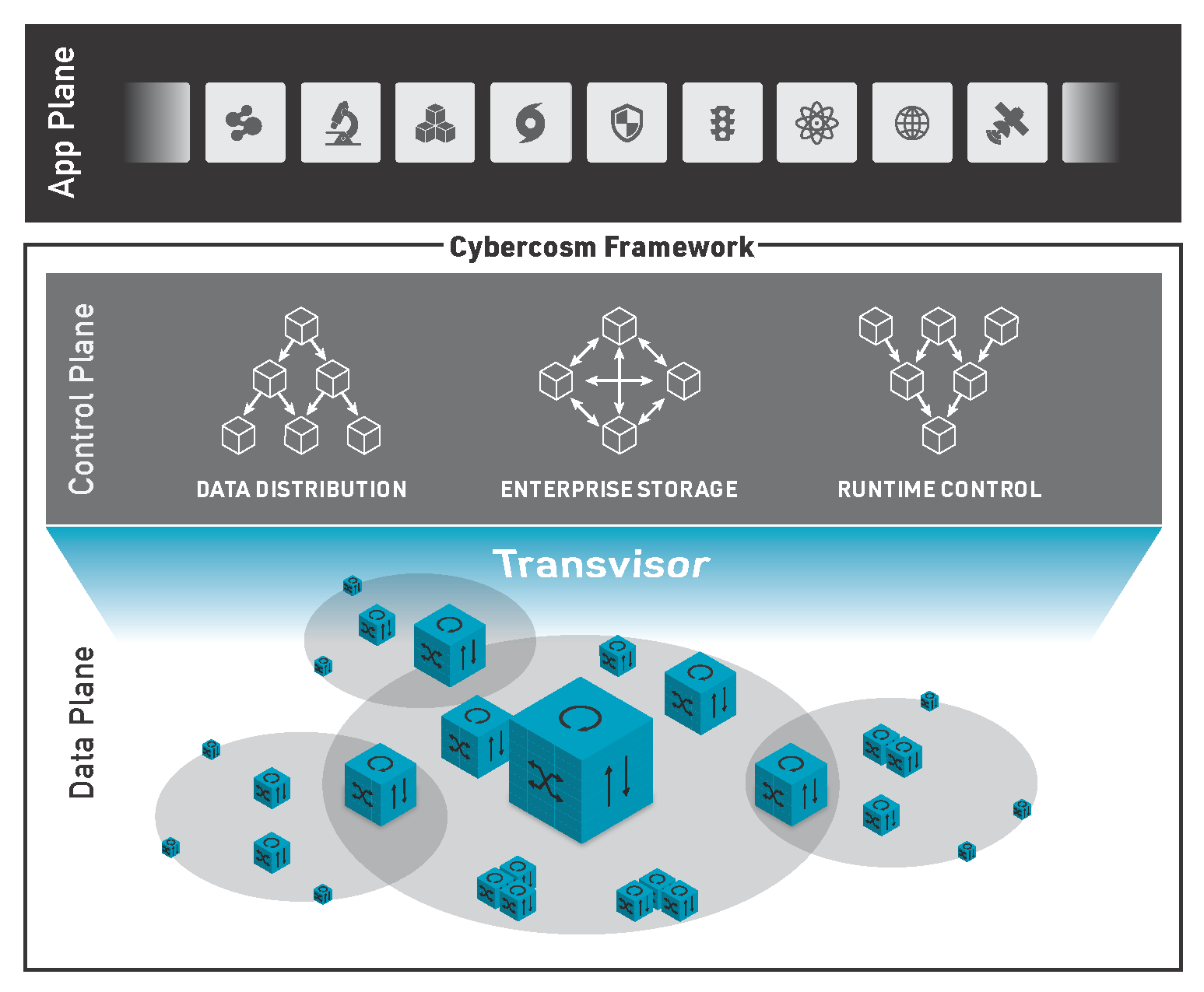}
\caption{\small High level overview of the 
architecture of an experimental \projectname implementation: the data layer composed of passive \platform nodes
accessed via the \emph{Internet Backplane Protocol} (IBP); the control layer
with diverse local area or global service managers (e.g., the Intelligent Data
Management System (IDMS), \lstore, and the Parallel Runtime Scheduling and
Execution Controller (PaRSEC); and the application layer control plane
managers used to perform work on data in the node plane. (See Section~\ref{software}).}
\label{f_planes}
\end{wrapfigure}

The \projectname project will create an experimental system that supports a distributed Federation of Resources (FoR) and develops the pieces needed to connect all components in the technology and application fabric. Figure~\ref{f_planes} describes the layers of the experimental system. The data plane is the lower layer and is the abstraction through which higher layers of a federated set of resources get access to local node and network resources. Services, such as globally routed packet delivery, process management, and file management are implemented in the control plane. The control plane takes responsibility for directing the forwarding of data within the data plane and directs the low level computational and storage functions of data plane nodes by invoking their node operating system. The data plane is responsible for the execution of simple and generic operations on buffers as issued by the data plane. Such operations can be limited in many dimensions (e.g., duration, input size, or memory utilization). It is the role of the control plane to implement sophisticated and specialized state management algorithms~\cite{beck2019interoperable}. The control and data planes can either co-reside on the same physical nodes, or they can be implemented separately. 
Thus, a centralized or less-distributed control plane service can potentially supervise the functions of a very large collection of 
well-connected data plane nodes. 

As noted above, \projectname's \platform layer does not simply provide technological tools; rather, it 
represents a rethinking of the architecture of data-intensive science application workflows by introducing a rigorous separation between the control plane and the data plane. 
Beyond simplifying system management and the deployment of new services, this extreme separation of control and data planes in the \projectname stack makes it possible for control plane services to share the metadata that defines a file or process without physically moving the underlying data objects or even the structural metadata itself, if it has been reflected into shareable data objects. We will use a widely applicable interchange format called the exNode~\cite{Bassi01MobileManagement} to define the movement of data between different high-level services that use or modify them.

\subsection{Initial Software Foundation}
\label{software}

The initial software foundation of the \projectname's data plane, sufficient to enable community experimentation, is the {\it Internet Backplane Protocol} (IBP) {\it depot}~\cite{Beck03anend-to-end}, which is a well tested overlay implementation of \projectname's exposed buffer architecture. Together with other low level topology and resource discovery operations, the IBP depot provides an interoperable buffer allocation, movement, and transformation service. Two mature control plane subsystems defined by data structures and protocols use the services of the data plane: {\it \lstore}~\cite{LStore} and {\it Intelligent Data Management Service (IDMS)}. \lstore implements a file system abstraction and accompanying enterprise storage services which implement both commodity/legacy and parallel high-performance access through new client interfaces. IDMS provides wide-area data distribution, replication, and transfer services under control of Flange, a flexible and expressive policy language~\cite{flange-icfc}. {\it PaRSEC} is a third control plane subsystem; it is a dataflow execution manager that describes parallel computations as program graphs and manages performance and fault tolerance policies among other necessary aspects of large-scale computing~\cite{PaRSEC}. PaRSEC is adapted to integrate elements of both \lstore and IDMS to extend its domain to data-intensive and compute-intensive workflows spanning our experimental system's interoperable wide-area data plane. The {\it Data Logistics Toolkit} or {\it DLT} ~\cite{dlt-web} (which includes the IBP depot, IDMS, Periscope, and other data plane elements), \lstore, and PaRSEC are all mature open-source projects, funded by the NSF, and the source codes are available for download from public repositories. In addition, complete site configurations of the DLT are available in containerized form for instantiation on standard infrastructure as a service (IaaS) environments.


\subsection{Core Functionalities, Capabilities, and Application-Defined Services}
\label{functionalities}

We expect our experimental system's agile computing service to have the following functionalities, capabilities, and application-defined services:

{\bf Control and Flexibility of Data Placement and Compute Placement:} In current HPC centers, big data must migrate to the center: large datasets become usable to scientists only after the data is at the HPC center and only when the scientists can run computing jobs on high-end resources. Even when the data is at the center, the data is not necessarily usable if a scientist cannot run a job due to, for example,  not having user accounts or an allocation. \projectname will provide a web-based workflow user interface (UI) so that a user can use a point-and-click graphical UI (GUI) to specify data's locations and/or paths of movement, as well as what kind of computing services need to be co-located at each location and along the path. 

{\bf Transparent, High-performance, and Fault-tolerant Data Transfer:}  Given a scientific workflow specification, there is initially a stage of preparing and testing the required computing services. To increase the flexibility users have in this phase, we will support users to port and install their own compute procedures as transvisor ops expressed and s the transformation of buffers. 
%
In \projectname, transparent realization of fault-tolerant and high-performance data transfer will be provided by logistical network mechanisms.

{\bf Flexible Control of Site Policy for Federation of Resources (FoR):} Non-uniform policies are one of the most challenging aspects of federating information and communication technology~(ICT) resources between organizations in a way that appears seamless to users. Many policies control access to data and allocation of ICT resources including storage, transfer, and computation. While mechanisms such as shared file systems and databases typically implement access control mechanisms that can, in principle, implement arbitrary higher level policies, in practice the mapping of institutional procedures and structures to these lower-level mechanisms can be cumbersome. 
%
\projectname will provide the same level of separation as condominium storage within shared object stores, but different storage allocations will be under the control of different control plane services under the administration of the local institution or site. Data will be shared between instances of a services at different sites under the control of policies defined by the control plane services, within parameters set by the local site and agreed to by the site. Without such a federation contract, data is by default isolated within the site, which can be as restricted as a single node (if so desired) by the local operator. 

{\bf Workflow Portability across the FoR:} One of the main benefits of \projectname's architecture is that an application scientist can view the computing environments on ``the edge'' and ``the center'' as implementing the same type, although differing in size, duration and stability. Underneath the uniform API, \projectname will transparently manage complexities, such as how to provide quality-of-service without requiring reservations, how to reduce application-level metrics like ``time to first result''  by bridging geographic distance, or how to find FoR resources that best fit a scientist's
needs~\cite{beckbeck1999qos}. 
\projectname's Resource Discovery is the main portal through which federated sites can register their resources.
When mapping a workflow onto FoR resources, the \projectname's resource discovery will serve as a light-weight global scheduler.

{\bf Maintenance of Data Integrity and Availability:} 
Access to critical operational data will be limited to system administration personnel over secure channels using best-practises methods and protocols (e.g., 2-factor authentication over SSL encrypted channels). Application data produced by or used by applications running on the \projectname architecture will be maintained as dictated by the level of criticality of the specific application.  Application users and maintainers will have the capability of implementing cross-grid redundancy, replication, encryption, authorization, and other access controls as required.

{\bf Inherent Support for Role-based, Federated Security:} The Cybercosm spanning layer is local to the depot and does not export a global service; services that use it cannot assume the global reachability of any given local node. By minimizing the ``target’’ that the node infrastructure offers to any external input or signal, Cybercosm can create a kind of ”white list” of allowable global services, as defined by the privileged control plane, leaving the node as impervious as possible to communication that is not part of such an authorized service. Then each higher level service can define its own strategy for authenticating, protecting and allowing access to the service it creates.

\paragraph*{\textit{Acknowledgements}:} This white paper builds on the results of earlier collaborative work. The authors want to thank the people who contributed to that effort: Ilkay Altintas, Srinivas Aluru, Laura Cadonati, Jian Huang, Tobin Issac, Piotr Luszczek, Greg Peterson, Aditya Prakash, Rampi Ramprasad, Paul Sheldon, D. Martin Swany, and Alan Tackett.

This work was initially funded---in the framework of the BDEC initiative---by the NSF (Grant No. 1849625), European Commission (Grant No. 800957) and DoE.

\paragraph*{\textit{Supplementary Materiel}:} A FAQ is available at \url{http://bit.ly/cybercosm_faq}. In it we will be adding and updating relevant explanations and material.